\documentclass[10pt,prl,nofootinbib,superscriptaddress,preprintnumbers,balancelastpage,twocolumn]{revtex4-1}
\pdfoutput=1
\usepackage{graphicx}
\usepackage{epstopdf}
\usepackage{bm}
\usepackage[utf8]{inputenc}
\usepackage{color}
\usepackage{soul}
\usepackage{amsmath}
\usepackage{amssymb}
\usepackage{mathtools}
\usepackage{xfrac}
\usepackage{rotating}
\usepackage{enumitem}
\usepackage{scrextend}
\usepackage{slashed}
\usepackage{axodraw2}
\usepackage{subfig}
\usepackage{array}

\newcommand{\be}{\begin{eqnarray}}
\newcommand{\ee}{\end{eqnarray}}

\newcommand{\eps}{\epsilon}
\newcommand{\epsv}{\varepsilon}
\newcommand{\Z}{\mathcal{Z}}
\newcommand{\T}{\mathcal{T}}

\usepackage{braket}
\usepackage{tcolorbox}
\usepackage[framemethod=tikz]{mdframed}

\newcommand{\MSbar}{{\ensuremath{\overline{\text{MS}}}}}

\chardef\MyArticleWithColor=\pdfcolorstackinit page direct{0 g}

\newcommand{\nn}{\nonumber}

\newcommand*\oline[1]{%
   \vbox{%
     \hrule height 0.5pt
     \kern0.5ex
     \hbox{%
       \kern-0.0em
       \ifmmode#1\else\ensuremath{#1}\fi
       \kern-0.0em
     }
   }
}

\def\trione#1#2#3
{
 \raisebox{-33pt}{
 \scalebox{0.7}{
 \begin{axopicture}{(90,80)(-45,-50)}
 \GluonArc(0,0)(30,30,180){2}{6}
 \GluonArc(0,0)(30,180,330){2}{6}
 \GluonArc(0,0)(30,-30,30){2}{2}
 \Gluon[double](-30,1)(-50,1){2}{1}
 \Gluon[double](25.98076212,15.)(43.30127020,25.){2}{1}
 \Gluon[double](25.98076212,-15.)(43.30127020,-25.){2}{1}
\CCirc(-30,0){3}{Black}{Red}
 \Vertex(25.98076212,15.){2}
 \Vertex(25.98076212,-15.){2}
\Text(-40,10){\Large $#1$ \Large}
\Text(40.30127020,33.){\Large$#2$ \Large}
\Text(40.30127020,-33.){\Large$#3$ \Large}
 \end{axopicture}
}}}

\def\trioneda
{
 \raisebox{-33pt}{
 \scalebox{0.7}{
 \begin{axopicture}{(70,80)(-30,-50)}
 \GluonArc(0,0)(30,30,180){2}{6}
 \GluonArc(0,0)(30,180,330){2}{6}
 \GluonArc(0,0)(30,-30,30){2}{2}
\CCirc(-30,0){4}{Black}{white}
\CCirc(-30,0){3}{Black}{Red}
\CCirc(25.98076212,15.){4}{Black}{white}
 \Vertex(25.98076212,15.){2}
\CCirc(25.98076212,-15.){4}{Black}{white}
 \Vertex(25.98076212,-15.){2}
 \Line(-24.,0.)(-36.,0.)
 \Line(-12.,-20.785)(-18.,-31.178)
 \Line(20.785,-12.)(31.178,-18.)
 \Text(-15.0,0.0){\large $p_1$ \large }
 \Text(-5.5000, -12.991){\large $p_1$ \large }
 \Text(14.991, -7.5000){\large$p_1$ \large}
 \end{axopicture}
}}}

\def\trionedb
{
 \raisebox{-33pt}{
 \scalebox{0.7}{
 \begin{axopicture}{(70,80)(-30,-50)}
 \GluonArc(0,0)(30,30,180){2}{6}
 \GluonArc(0,0)(30,180,330){2}{6}
 \GluonArc(0,0)(30,-30,30){2}{2}
\CCirc(-30,0){4}{Black}{white}
\CCirc(-30,0){3}{Black}{Red}
\CCirc(25.98076212,15.){4}{Black}{white}
 \Vertex(25.98076212,15.){2}
\CCirc(25.98076212,-15.){4}{Black}{white}
 \Vertex(25.98076212,-15.){2}
 \Line(-24.,0.)(-36.,0.)
 \Line(-12.,20.785)(-18.,31.178)
 \Line(20.785,12.)(31.178,18.)
 \Text(-15.0,0.0){\large $p_1$ \large }
 \Text(-5.5000, 12.991){\large $p_2$ \large }
 \Text(14.991, 7.5000){\large$p_2$ \large}
 \end{axopicture}
}}}

\def\trionedc
{
 \raisebox{-33pt}{
 \scalebox{0.7}{
 \begin{axopicture}{(70,80)(-30,-50)}
 \GluonArc(0,0)(30,30,180){2}{6}
 \GluonArc(0,0)(30,180,330){2}{6}
 \GluonArc(0,0)(30,-30,30){2}{2}
\CCirc(-30,0){4}{Black}{white}
\CCirc(-30,0){3}{Black}{Red}
\CCirc(25.98076212,15.){4}{Black}{white}
 \Vertex(25.98076212,15.){2}
\CCirc(25.98076212,-15.){4}{Black}{white}
 \Vertex(25.98076212,-15.){2}
 \Line(-23.967,1.2561)(-35.951,1.7795)
 \Line(-24.,0.)(-36.,0.)
 \Line(-23.967,-1.2561)(-35.951,-1.7795)
 \Text(-15.0,0.0){\large $p_1$ \large }
 \end{axopicture}
}}}

\def\trionedcm
{
 \raisebox{-33pt}{
 \scalebox{0.7}{
 \begin{axopicture}{(70,80)(-30,-50)}
 \GluonArc[width=1.5](0,0)(30,30,180){2}{6}
 \GluonArc(0,0)(30,180,330){2}{6}
 \GluonArc(0,0)(30,-30,30){2}{2}
\CCirc(-30,0){4}{Black}{white}
\CCirc(-30,0){3}{Black}{Red}
\CCirc(25.98076212,15.){4}{Black}{white}
 \Vertex(25.98076212,15.){2}
\CCirc(25.98076212,-15.){4}{Black}{white}
 \Vertex(25.98076212,-15.){2}
 \Line(-23.967,1.2561)(-35.951,1.7795)
 \Line(-24.,0.)(-36.,0.)
 \Line(-23.967,-1.2561)(-35.951,-1.7795)
 \Text(-15.0,0.0){\large $p_1$ \large }
 \end{axopicture}
}}}

\def\trionedcmb
{
 \raisebox{-33pt}{
 \scalebox{0.7}{
 \begin{axopicture}{(70,80)(-30,-50)}
 \GluonArc(0,0)(30,30,180){2}{6}
 \GluonArc(0,0)(30,180,330){2}{6}
 \GluonArc[width=1.5](0,0)(30,-30,30){2}{2}
\CCirc(-30,0){4}{Black}{white}
\CCirc(-30,0){3}{Black}{Red}
\CCirc(25.98076212,15.){4}{Black}{white}
 \Vertex(25.98076212,15.){2}
\CCirc(25.98076212,-15.){4}{Black}{white}
 \Vertex(25.98076212,-15.){2}
 \Line(-23.967,1.2561)(-35.951,1.7795)
 \Line(-24.,0.)(-36.,0.)
 \Line(-23.967,-1.2561)(-35.951,-1.7795)
 \Text(-15.0,0.0){\large $p_1$ \large }
 \end{axopicture}
}}}

\def\trionedcIR
{
 \raisebox{-33pt}{
 \scalebox{0.7}{
 \begin{axopicture}{(70,80)(-30,-50)}
 \GluonArc(0,0)(30,0,180){2}{6}
 \GluonArc(0,0)(30,180,360){2}{6}
\CCirc(-30,0){4}{Black}{white}
\CCirc(-30,0){3}{Black}{Red}
\CCirc(30,0){3}{Black}{white}
 \Line(-23.967,1.2561)(-35.951,1.7795)
 \Line(-24.,0.)(-36.,0.)
 \Line(-23.967,-1.2561)(-35.951,-1.7795)
 \Text(-15.0,0.0){\large $p_1$ \large }
 \end{axopicture}
}}}

\def\trionedcRE
{
 \raisebox{-33pt}{
 \scalebox{0.7}{
 \begin{axopicture}{(20,80)(20,-50)}
 \GluonArc[width=1.5](0,0)(30,-30,30){2}{2}
\CCirc(25.98076212,15.){5}{Black}{white}
\CCirc(25.98076212,15.){4}{Black}{white}
 \Vertex(25.98076212,15.){2}
\CCirc(25.98076212,-15.){5}{Black}{white}
\CCirc(25.98076212,-15.){4}{Black}{white}
 \Vertex(25.98076212,-15.){2}
 \end{axopicture}
}}}

\def\trionedcIRUV
{
 \raisebox{-33pt}{
 \scalebox{0.7}{
 \begin{axopicture}{(70,80)(-30,-50)}
 \GluonArc[width=1.5](0,0)(30,0,180){2}{6}
 \GluonArc(0,0)(30,180,360){2}{6}
\CCirc(-30,0){4}{Black}{white}
\CCirc(-30,0){3}{Black}{Red}
\CCirc(30,0){3}{Black}{white}
 \Line(-23.967,1.2561)(-35.951,1.7795)
 \Line(-24.,0.)(-36.,0.)
 \Line(-23.967,-1.2561)(-35.951,-1.7795)
 \Text(-15.0,0.0){\large $p_1$ \large }
 \end{axopicture}
}}}

\def\tritwoM
{
 \raisebox{-33pt}{
 \scalebox{0.7}{
 \begin{axopicture}{(70,80)(-30,-50)}
 \GluonArc(0,0)(30,30,90){2}{2}
 \GluonArc[width=1.5](0,0)(30,90,180){2}{4}
 \GluonArc(0,0)(30,180,270){2}{4}
 \GluonArc(0,0)(30,270,330){2}{2}
 \GluonArc(0,0)(30,-30,30){2}{2}
 \Gluon(0,30)(0,-30){2}{5}
\Vertex(0,30){2}
\Vertex(0,-30){2}
 \CCirc(-30,0){4}{Black}{white}
 \CCirc(25.98076212,15.){4}{Black}{white}
 \CCirc(25.98076212,-15.){4}{Black}{white}
\CCirc(-30,0){3}{Black}{Red}
 \Vertex(25.98076212,15.){2}
 \Vertex(25.98076212,-15.){2}
  \Line(-23.967,1.2561)(-35.951,1.7795)
 \Line(-24.,0.)(-36.,0.)
 \Line(-23.967,-1.2561)(-35.951,-1.7795)
 \Text(-15.0,0.0){\large $p_1$ \large }
 \end{axopicture}
 }}}

 \def\tritwoMuva
{
 \raisebox{-33pt}{
 \scalebox{0.7}{
 \begin{axopicture}{(40,80)(-30,-50)}
 \GluonArc[width=1.5](0,0)(30,90,180){2}{4}
 \GluonArc(0,0)(30,180,270){2}{4}
 \Gluon(0,30)(0,-30){2}{5}
 \CCirc(0,30){4}{Black}{white}
\Vertex(0,30){2}
 \CCirc(0,-30){4}{Black}{white}
\Vertex(0,-30){2}
 \CCirc(-30,0){4}{Black}{white}
\CCirc(-30,0){3}{Black}{Red}
  \Line(-23.967,1.2561)(-35.951,1.7795)
 \Line(-24.,0.)(-36.,0.)
 \Line(-23.967,-1.2561)(-35.951,-1.7795)
 \Text(-15.0,0.0){\large $p_1$ \large }
 \end{axopicture}
 }}}

\def\tritwoMira
{
 \raisebox{-33pt}{
 \scalebox{0.7}{
 \begin{axopicture}{(70,80)(-30,-50)}
 \GluonArc(0,0)(30,30,180){2}{6}
 \GluonArc(0,0)(30,180,330){2}{6}
 \GluonArc(0,0)(30,-30,30){2}{2}
 \CCirc(-30,0){3}{Black}{white}
 \CCirc(25.98076212,15.){4}{Black}{white}
 \CCirc(25.98076212,-15.){4}{Black}{white}
 \Vertex(25.98076212,15.){2}
 \Vertex(25.98076212,-15.){2}
 \end{axopicture}
}}}

\def\tritwouvb
{
 \raisebox{-33pt}{
 \scalebox{0.7}{
 \begin{axopicture}{(40,80)(0,-50)}
 \GluonArc(0,0)(30,30,90){2}{2}
 \GluonArc(0,0)(30,270,330){2}{2}
 \GluonArc(0,0)(30,-30,30){2}{2}
 \Gluon(0,30)(0,-30){2}{5}
 \CCirc(0,30){4}{Black}{white}
 \CCirc(0,-30){4}{Black}{white}
\Vertex(0,30){2}
\Vertex(0,-30){2}
 \CCirc(25.98076212,15.){4}{Black}{white}
 \CCirc(25.98076212,-15.){4}{Black}{white}
 \Vertex(25.98076212,15.){2}
 \Vertex(25.98076212,-15.){2}
 \end{axopicture}
 }}}

\def\tritwouvbrem
{
 \raisebox{-33pt}{
 \scalebox{0.7}{
 \begin{axopicture}{(70,80)(-30,-50)}
 \GluonArc[width=1.5](0,0)(30,0,180){2}{6}
 \GluonArc(0,0)(30,180,360){2}{6}
 \CCirc(30,0){4}{Black}{white}
 \CCirc(-30,0){4}{Black}{white}
\CCirc(-30,0){3}{Black}{Red}
  \Line(-23.967,1.2561)(-35.951,1.7795)
 \Line(-24.,0.)(-36.,0.)
 \Line(-23.967,-1.2561)(-35.951,-1.7795)
 \Text(-15.0,0.0){\large $p_1$ \large }
 \end{axopicture}
 }}}

\def\dvertex
{
 \begin{axopicture}{(15,5)(-2.5,-2.5)}
    \Vertex(0,0){1.5}  
    \Line(-5.,0.)(5.,0.)   
    \Text(7.5,0){$p$}
\end{axopicture}
}

\def\nvertex
{
 \begin{axopicture}{(5,5)(-2.5,-2.5)}
\CCirc(0,0){3}{Black}{white}
 \Vertex(0,0){1.5}
\end{axopicture}
}

\begin{document}
\preprint{RBRC-1318}
\preprint{Nikhef 2019-033}
\graphicspath{{Figures/}}

\title{
Two- and three-loop anomalous dimensions of Weinberg's dimension-six CP-odd gluonic operator
}

\author{Jordy de Vries}
\affiliation{Amherst Center for Fundamental Interactions, Department of Physics, University of Massachusetts, Amherst, MA 01003}
\affiliation{RIKEN BNL Research Center, Brookhaven National Laboratory,
Upton, New York 11973-5000, USA}

\author{Giulio Falcioni} \affiliation{Higgs Centre for Theoretical Physics, School of Physics and Astronomy, The University of Edinburgh, Edinburgh EH9 3FD, Scotland, UK}

\author{Franz Herzog}
\affiliation{Nikhef Theory Group, Science Park 105, 1098 XG Amsterdam, The Netherlands}

\author{Ben Ruijl}
\affiliation{ETH Z\"urich,
R\"amistrasse 101, %
8092 Z\"urich, Switzerland}

\begin{abstract}
We apply a fully automated extension of the $R^*$-operation capable of calculating higher-loop anomalous dimensions of n-point Green's functions of arbitrary, possibly non-renormalisable, local Quantum Field Theories. We focus on the case of the CP-violating Weinberg operator of the Standard Model Effective Field Theory whose anomalous dimension is so far known only at one loop. We calculate the two-loop anomalous dimension in full QCD and the three-loop anomalous dimensions in the limit of pure Yang-Mills theory. We find sizeable two-loop and large three-loop corrections, due to the appearance of a new quartic group invariant. We discuss phenomenological implications for electric dipole moments and future applications of the method.
\end{abstract}

\date{\today}

\maketitle

\section{Introduction}
The absence of evidence for beyond-the-Standard Model (BSM) physics in high-energy proton-proton collisions at the LHC, and in large classes of low-energy precision measurements all indicate that the scale of BSM physics ($\Lambda$) is significantly higher than the electroweak scale ($v$): $\Lambda \gg v \simeq 246$ GeV. In such a scenario, the effects of BSM physics at low energies $E \ll \Lambda$, can be described in terms of effective operators consisting of SM fields that obey the Standard Model (SM) gauge and Lorentz symmetries~\cite{Buchmuller:1985jz,Grzadkowski:2010es,Brivio:2017vri}. The resulting framework is called the SM effective field theory (SMEFT). The SMEFT Lagrangian contains an infinite number of operators that can be ordered by their dimension. Effects of higher-dimensional operators on low-energy observables are suppressed by additional powers of $E/\Lambda$.

The connection between observables at (relatively) low energies and the SMEFT operators at the scale $\Lambda$, where the effective operators can be matched to specific UV-complete BSM models, is determined by renormalisation-group equations (RGEs). The RGEs depend on anomalous dimensions that can be calculated in perturbation theory in an expansion in small coupling constants. The complete one-loop anomalous dimension matrix of dimension-six SMEFT operators has been obtained \cite{Jenkins:2013zja,Jenkins:2013wua,Alonso:2013hga}. Already for dimension-six, the number of operators is large and the general mixing structure of the RGEs is rather complex. It has been observed that the one-loop anomalous dimension matrix is almost holomorphic  \cite{Alonso:2014rga}, but it is not clear whether this feature extends to higher order. Higher-order anomalous dimensions have been calculated for subsets of dimension-six operators \cite{Altarelli:1980fi,Buras:1989xd,Buras:1991jm,Ciuchini:1992tj,Ciuchini:1993vr,Misiak:1994zw,Chetyrkin:1996vx,Chetyrkin:1997fm,Ciuchini:1998ix,Buras:2000if,Gracey:2000am,Gambino:2003zm,Gorbahn:2004my,Gorbahn:2005sa,Degrassi:2005zd, Czakon:2006ss}, but due to the hard technical nature of the calculations the complete matrix is not known. Higher-order anomalous dimensions can be used to (1) improve the precision of SMEFT contributions to LHC processes or low-energy precision observables, (2) to study the convergence of the perturbative expansions, and (3) to investigate the structure of the SMEFT mixing pattern.

In this paper we extend the $R^*$-operation to the framework of the Standard Model Effective Field Theory (SMEFT), and develop an efficient and highly automated method to calculate higher-order QCD anomalous dimensions of SMEFT operators. The $R^*$-operation provides a way to subtract UV and IR divergences from Euclidean Feynman diagrams, taking care of the combinatorics of overlapping divergences~\cite{Chetyrkin:1982nn,Chetyrkin:2017ppe,Chetyrkin:1984xa}. Recently, it has been extended to Feynman diagrams with arbitrary numerator structure~\cite{Herzog:2017bjx}. So far the $R^*$-operation has been used extensively in calculations of anomalous dimensions in QCD; see, e.g., \cite{Baikov:2016tgj,Herzog:2017ohr,Herzog:2017dtz,Chetyrkin:2017bjc}. However, the $R^*$-method is not limited to pure QCD and can be applied to arbitrary local quantum field theories.

To avoid the complicated mixing structure of SMEFT operators, we focus on a specific SMEFT dimension-six operator. Once tested and developed, the method can be extended to a larger set of operators without too many additional complications. The operator we now consider is the CP-violating gluonic operator, often called ``the Weinberg operator''~\cite{Weinberg:1989dx}, defined as
\begin{equation}
  \mathcal L_{6} =\frac{ C_W}{6}f^{abc} \epsilon^{\mu\nu\alpha\beta}G^a_{\alpha\beta}G^b_{\mu\rho}G^{c\,\rho}_{\nu}\equiv C_W\,O_W\,,
  \label{def:weinberg}
\end{equation}
in terms of the gluon field strength $G^a_{\alpha\beta}$, the Levi-Civita tensor (LCT) $\epsilon^{\mu\nu\alpha\beta}$, the gauge group structure constants $f^{abc}$, and the Wilson coefficient $C_W \sim 1/\Lambda^2$. The Weinberg operator is induced in various classes of BSM models with additional CP-violating phases such as supersymmetric models, two-Higgs doublet models, and models with leptoquarks \cite{Demir:2002gg,Dekens:2014jka,Abe:2017sam}. The Weinberg operator is also induced from dimension-six CP-odd operators in the SMEFT Lagrangian involving heavy quarks, prominent examples being heavy-quark chromo-electric dipole moments \cite{Chien:2015xha,Cirigliano:2016nyn}, and heavy-quark Yukawa interactions \cite{Brod:2013cka,Chien:2015xha,Brod:2018pli}. At lower energies, the Weinberg operator leads to nonzero electric dipole moments (EDMs) of nucleons, nuclei, and diamagnetic atoms (such as ${}^{199}$Hg \cite{Graner:2016ses}). Current experimental EDM limits \cite{Chupp:2017rkp} set strong constraints on BSM models that induce the Weinberg operator. 

To use EDM limits to constrain the Weinberg operator, and the associated BSM models, it is necessary to evolve the Weinberg operator from the high-energy scale where it is induced to the low-energy scale where the Weinberg operator is matched to hadronic CP-violating operators. This evolution is determined by the anomalous dimension of the Weinberg operator. 
The one-loop anomalous dimension of the Weinberg operator was obtained in the original paper by Weinberg~\cite{Weinberg:1989dx}, albeit with the wrong sign, finding sizeable QCD corrections from the evolution of $C_W$ from high- to low-energy scales. The calculation was corrected in Refs.~\cite{BraatenPRL,Braaten:1990zt,Chang:1990dja} that also calculated the mixing, proportional to the small quark masses, of the Weinberg operator into the quark chromo-electric dipole moment. The only other operator $C_W$ can mix with is the QCD theta term $\sim \epsilon^{\mu\nu\alpha\beta}G^a_{\alpha\beta}G^a_{\mu\nu}$, but this mixing is of little phenomenological use as the bare theta term is an unknown SM parameter. Furthermore, the renormalised theta term vanishes after a Peccei-Quinn mechanism \cite{Peccei:1977hh}. We will not consider the mixing into the theta term in this paper.

The potential phenomenological implications of the unknown higher-loop anomalous dimensions of $C_W$, in addition to the high complexity of the Feynman rules induced by the Weinberg operator, make determining the higher-order corrections a suitable real-world test case for the $R^*$-method.

\section{The background field method}

The renormalisation of Green's functions with a single insertion 
of the Weinberg operator $O_W$ requires in general a counterterm matrix
$Z_{ij}$. The matrix $Z_{ij}$ takes into account mixing 
with all the operators of equal or smaller mass dimension, which share the same 
quantum numbers as $O_W$~\cite{Itzykson:1980rh,Pascual:1984zb}:
\begin{equation}
  O_W^R = \sum_j Z_{Wj}O_j^B \,.
  \label{def:OWR}
\end{equation}
The operators that contribute to eq. \eqref{def:OWR} are divided into
three classes~\cite{Dixon:1974ss,Joglekar:1975nu}:
\begin{itemize}
  \item gauge invariant (GI) {\textit{physical operators}}, 
  \item operators that vanish after applying the classical equations of
    motion ({\textit{eom}}),
  \item BRST-exact operators.
\end{itemize}
The {\textit{eom}} and the BRST-exact operators are unphysical, since
they have vanishing $S$-matrix elements. Nevertheless they have non-zero
Green's functions and associated UV counterterms which mix with $O_W$,
as in eq.~\eqref{def:OWR}.

We use the background field method
\cite{tHooft:1975uxh,DeWitt:1980jv,Abbott:1980hw,Abbott:1981ke} to
simplify the mixing pattern of eq. \eqref{def:OWR}. The main advantage
of this method is that it preserves gauge invariance of the background
field, so that the UV counterterms of 1PI correlators of the background fields involve only GI operators. This feature puts strong constraints on the non-physical operators in eq. (\eqref{def:OWR}): BRST-exact operators will not contribute to the UV counterterms and gauge invariance will restrict {\textit{eom}} operators too. For example, in Yang-Mills theory without fermions ($n_f=0$), there is only one independent gauge invariant {\textit{eom}} operator
\begin{equation}
  \widetilde{O}_E=\frac{1}{4}\epsilon^{\mu_1\mu_2\mu_3\mu_4}\left(D_{\mu_1}G_{\mu_2\mu_3}\right)^a\left(D^\lambda G_{\lambda\mu_4}\right)^a \,,
  \label{eq:EOMGI}
\end{equation}
where $D^a_\mu$ is the covariant derivative. However, $\widetilde{O}_E$ vanishes by the Bianchi identity. Therefore it is impossible to construct purely gluonic unphysical operators mixing with $O_W$. In the rest of this paper, we will compute directly the UV counterterm $Z_{WW}$, which cancels the local UV divergences of the background field correlators with a single $O_W$ vertex. The $R^*$-operation, which
subtracts recursively all the UV subdivergences of the diagrams in 
a fully automated way, will be the key tool to isolate the gauge invariant local UV divergence 
of the correlators.
We calculate $Z_{WW}$ at the two-loop level in full QCD. At three loops, we only
consider $Z_{WW}$ in Yang-Mills theory without fermions. The corresponding diagrams are the most computationally demanding. In general, the
inclusion of fermions will also generate off-diagonal mixing $Z_{Wj}$ with quark
(chromo)-electric operators \cite{Dekens:2013zca,Cirigliano:2020msr}. These contributions will be the subject of a separate
work.

\section{The $R^*$-operation for general Feynman diagrams}
\label{sec:rstar}
\begin{figure}
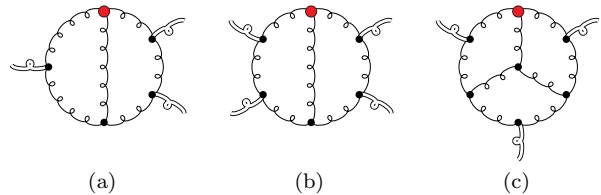

\centering
\subfloat[]{{
 \raisebox{-26pt}{
 \scalebox{0.7}{
 \begin{axopicture}{(100,80)(-45,-50)}
 \GluonArc(0,0)(30,30,90){2}{2}
 \GluonArc(0,0)(30,90,180){2}{4}
 \GluonArc(0,0)(30,180,270){2}{4}
 \GluonArc(0,0)(30,270,330){2}{2}
 \GluonArc(0,0)(30,-30,30){2}{2}
 \Gluon[double](-30,1)(-50,1){2}{1}
 \Gluon(0,30)(0,-30){2}{5}
 \Gluon[double](25.98076212,15.)(43.30127020,25.){2}{1}
 \Gluon[double](25.98076212,-15.)(43.30127020,-25.){2}{1}
\CCirc(0,30){3}{Black}{Red}
 \Vertex(-30,0){2}
 \Vertex(0,-30){2}
 \Vertex(25.98076212,15.){2}
 \Vertex(25.98076212,-15.){2}
 \end{axopicture}
 }}}}
\subfloat[]{{
\raisebox{-26pt}{
 \scalebox{0.7}{
 \begin{axopicture}{(100,80)(-45,-50)}
 \GluonArc(0,0)(30,30,90){2}{2}
 \GluonArc(0,0)(30,270,330){2}{2}
 \GluonArc(0,0)(30,90,150){2}{2}
 \GluonArc(0,0)(30,150,210){2}{2}
 \GluonArc(0,0)(30,210,270){2}{2}
\GluonArc(0,0)(30,-30,30){2}{2}
 \Gluon(0,30)(0,-30){2}{5}
 \Gluon[double](25.98076212,15.)(43.30127020,25.){2}{1}
 \Gluon[double](25.98076212,-15.)(43.30127020,-25.){2}{1}
 \Gluon[double](-25.98076212,15.)(-43.30127020,25.){2}{1}
 \Gluon[double](-25.98076212,-15.)(-43.30127020,-25.){2}{1}
\CCirc(0,30){3}{Black}{Red}
 \Vertex(-25.98076212,15.){2}
 \Vertex(-25.98076212,-15.){2}
 \Vertex(0,-30){2}
 \Vertex(25.98076212,15.){2}
 \Vertex(25.98076212,-15.){2}
 \end{axopicture}
 }}}}
\subfloat[]{{
 \raisebox{-26pt}{
 \scalebox{0.7}{
  \begin{axopicture}{(100,80)(-45,-50)}
 \GluonArc(0,0)(30,30,90){2}{2}
 \GluonArc(0,0)(30,270,330){2}{2}
 \GluonArc(0,0)(30,90,150){2}{2}
 \GluonArc(0,0)(30,150,210){2}{2}
 \GluonArc(0,0)(30,210,270){2}{2}
\GluonArc(0,0)(30,-30,30){2}{2}
 \Gluon(0,30)(0,0){2}{2}
 \Gluon(25.98076212,-15.)(0,0){2}{2}
 \Gluon(-25.98076212,-15.)(0,0){2}{2}
 \Gluon[double](25.98076212,15.)(43.30127020,25.){2}{1}
 \Gluon[double](-25.98076212,15.)(-43.30127020,25.){2}{1}
 \Gluon[double](0,-30)(0,-50){2}{1}
\CCirc(0,30){3}{Black}{Red}
 \Vertex(-25.98076212,15.){2}
 \Vertex(-25.98076212,-15.){2}
 \Vertex(0,-30){2}
 \Vertex(25.98076212,15.){2}
 \Vertex(25.98076212,-15.){2}
\Vertex(0,0){2}
 \end{axopicture}
 }}}}
 \caption{Two-loop diagrams with three (a) and four (b) external background 
 fields and a three-loop diagram with three external background fields (c). Gluons are denoted by curly lines and an insertion of the Weinberg operator by a red dot.}
 \label{fig:diags}
\end{figure}

The renormalisation constant $Z_{WW}$ can be extracted from the 1PI correlator $\Gamma^{nb}_{W}$ of $n$ background 
fields with a single insertion of $O_W$ --- see figure \ref{fig:diags} for examples of Feynman diagrams --- by acting on it 
with the UV-counterterm operation $\Z$:
\begin{align}
  Z_{WW}Z_b^{3/2} &O_W^{nb}(C_W;p_1,\ldots,p_n)=\nonumber\\
  &\Z\Big(\Gamma^{nb}_{W}(\alpha_s,C_W;p_1,\ldots,p_n)\Big)\,,
\end{align}
where $Z_b$ is the well-known wave function renormalisation of the background gauge field \cite{Abbott:1980hw,Abbott:1981ke} and
the $\Z$-operation is defined to include counterterms for all the UV-subdivergences of $\Gamma^{nb}_{W}$. 
Here $O_W^{nb}$ denotes the Feynman rule of the $n$ background-field vertex generated by the Weinberg operator. 
The calculation of renormalisation constants can be simplified by nullifying the external momenta of $\Gamma^{nb}_{W}$ after 
applying a Taylor expansion in the external momenta whose order equals the superficial degree of divergence of the correlator; 
for the case of $\Gamma^{nb}_{W}$ this is $\omega(\Gamma^{nb}_W)=6-n$, e.g. $\omega(\Gamma^{3b}_W)=3$. 
After the external momenta are nullified, a convenient scale can be reintroduced into the correlator by introducing either arbitrarily chosen 
external momenta into each Feynman diagram or inserting a mass into a single propagator. This is known as the procedure of infrared rearrangement (IRR)~\cite{Vladimirov:1979zm}. Nullifying the external momenta introduces new IR-divergences, which are fully automatically subtracted by the local $R^*$-operation.

In contrast to earlier works which made use of the $R^*$-operation, we now act it on Feynman diagrams before contracting any of the Feynman rules. This makes the algorithm more efficient and allows us to directly compute the UV counterterm of a particular diagram in \MSbar{}. As discussed in Ref.~\cite{Herzog:2017bjx}, this is not possible if the Feynman rules are contracted before the action of $R^*$. The algorithm of \cite{Herzog:2017bjx} was therefore only capable of extracting the pole terms of self-energy diagrams from simpler self-energy diagrams. The more general method used in this work allows us to reduce the calculation of UV-counterterms of correlators with arbitrary numbers of external legs to the calculation of massless self-energy diagrams of one loop less. Therefore, this method has quite some advantages compared to the former, but its implementation introduces new complications. For instance, the Taylor expansion must  be applied before the contraction of the Feynman rules, which leads to a new set of `differentiated' Feynman rules (essentially new vertices and propagators). A more  detailed overview of the method will be given in an upcoming publication \cite{TBPBF}.
Employing this new formalism we write
\be
\label{eq:Zrelation}
\Z\Big(\Gamma^{nb}_{W}\Big)=-K \bar R^* \Big(\T_{p_1,\ldots,p_n}^{(6-n)}\Gamma^{nb}_{W}\Big|_{p_i=0}\Big)\,,
\ee
where the operation $\bar R^*$  acting on a Feynman diagram $\Gamma$ subtracts from it counterterms for all UV subdivergences and all IR divergences.  The operation $K$ extracts the single- and multi-pole contributions in the dimensional regulator $\eps=(4-D)/2$ and $\T_{p_1,\ldots,p_n}^{(\omega)}$ denotes the Taylor expansion operator for the order $\omega$-term in the expansion around the (external) momenta $p_1,\ldots,p_n$. More precisely (although we leave the nitty gritty of the graph combinatorics to the literature; see, e.g., \cite{Herzog:2017bjx} or \cite{Beekveldt:2020kzk}), the $\bar R^*$ is defined as:
\be
\bar R^*(\Gamma)=\sum_{\gamma\cap \tilde \gamma=\emptyset} \tilde \Z(\tilde \gamma)\ast\Z(\gamma)\ast\Gamma\setminus\tilde\gamma/\gamma\,,
\ee
where the sum goes overall non-intersecting UV-subgraphs $\gamma$ (but not including the full graph $\Gamma$) and IR-subgraphs $\tilde\gamma$ (including also $\Gamma$ in the case that $\Gamma$ is a log-divergent vacuum graph). The operation $\tilde\Z$ here is the IR-counterterm operation which can always be rewritten in terms of the UV-counterterm operation of $\tilde \gamma$ and its subgraphs. The remaining contracted graph $\Gamma\setminus\tilde\gamma/\gamma$ is constructed by deleting the IR subgraph and then contracting the UV subgraph to a point. The $*$-operation denotes insertion of the counterterms into the remaining contracted graph; it reduces to the usual multiplication for log-divergent counterterms.

To simplify the calculation of the counterterm we use the fact that the $\Z$-operation commutes with the Taylor expansion operator. In the following we give some examples of this procedure. Let us start with the calculation of the UV-counterterm of the following one-loop graph:
\begin{equation}
\Z\Big(\trione{p_1}{p_2}{p_3} \Big)=\Z\Big( \T^{(3)}_{p_1,p_2}\trione{p_1}{p_2}{-p_{12}} \Big)
\end{equation}
Here the thicker red vertex continues to denote the Weinberg operator. The Taylor expansion operator has to applied before the contraction of the Feynman rules. This procedure can be carried out diagrammatically, after choosing a momentum routing. Here we pick $p_1$ and $p_2$ as independent external momenta and write $p_3=-p_1-p_2=-p_{12}$. To simplify the derivative calculation it is usually best to pick the shortest paths through the diagram. Subsequently one differentiates along the path using nothing but the product and chain rule; thereby one generates a (potentially large) sum of new diagrams which contain differentiated propagators and vertices, and with the external momenta nullified. 

We graphically depict the differentiated vertex or Feynman rule with a small line $\dvertex$, and a label denoting that the momentum $p$ has been differentiated with operator $p.\partial_p$. The orientation of the derivative fixes the direction of the path. Subsequently we need to nullify the external momenta; we introduce an encircled vertex $\nvertex$ to denote that the external momentum (in  our case always associated to the background field) has been nullified. Naturally both nullification and derivation will also occur simultaneously on certain vertices. In this case the expression is to be understood as $p.\partial_p(\bullet)|_{p=0}$. After the derivations, external momentum nullifications and mass insertions have been carried out we then obtain (writing just a few terms):
\begin{align}
\label{eq:oneloopex}
&\Z\Big(\T^{(3)}_{p_1,p_2}\trione{p_1}{p_2}{p_3} \Big)=+\frac{1}{3!}\Z\Big(\trionedc\Big)\nn\\
&\qquad +\frac{1}{3!}\Z\Big( \trioneda \Big) +\frac{1}{2!}\Z\Big(\trionedb\Big)\\
&\qquad +\text{many other terms}\nn
\end{align}
To evaluate one of the counterterms we need to introduce a scale back into the vacuum diagrams. Given that they are logarithmically divergent, the counterterms are independent of all scales. A simple choice is to insert a mass into a single propagator; denoted graphically by a thicker line. When possible it is convenient to insert the mass in such a way as to prevent IR-divergences from occuring. Let us consider just the first term on the right hand side of eq. (\ref{eq:oneloopex}). A possibile IR rearrangement to avoid the IR-divergence is:
\begin{equation}
  \label{eq:oneloopNoIR}
\Z\Big(\trionedc\Big)=-K\Big(\trionedcm\Big)\,.
\end{equation}

The absence of infrared singularities in the rearranged diagram of eq.~(\ref{eq:oneloopNoIR}) is verified by applying a power counting procedure in the infrared region of the loop momentum, characterised by $k^\mu \rightarrow \lambda k^\mu$ with $\lambda\ll 1$. In this limit, each massless propagator diverges as ${\cal{O}}\left(\frac{1}{\lambda^2}\right)$. However, each vertex provides a suppression in the numerator, given respectively by
\begin{equation}
\begin{array}{cc}
\raisebox{-33pt}{
 \scalebox{0.7}{
 \begin{axopicture}{(80,80)(-30,-50)}
 \GluonArc(0,0)(30,120,180){2}{4}
 \GluonArc(0,0)(30,180,240){2}{4}
\CCirc(-30,0){4}{Black}{white}
\CCirc(-30,0){3}{Black}{Red}
 \Line(-23.967,1.2561)(-35.951,1.7795)
 \Line(-24.,0.)(-36.,0.)
 \Line(-23.967,-1.2561)(-35.951,-1.7795)
 \PText(-14,0)(0)[l]{$\simeq{\cal{O}}(1),$}
 \end{axopicture}
}}
&
\raisebox{-33pt}{
 \scalebox{0.7}{
 \begin{axopicture}{(80,80)(-30,-50)}
 \GluonArc(0,0)(30,120,180){2}{4}
 \GluonArc(0,0)(30,180,240){2}{4}
\CCirc(-30,0){4}{Black}{white}
\Vertex(-30,0){2}
\PText(-14,0)(0)[l]{$\simeq{\cal{O}}(\lambda).$}
 \end{axopicture}
}}
\end{array}
\end{equation}
Therefore, the infrared region gives a vanishing contribution to the diagram
\begin{equation}
  \trionedcm\simeq O\left(\lambda^2\right).
\end{equation}

An IR-counterterm would have been required had we used instead:
\begin{align}
&\Z\Big(\trionedc\Big)=-K\Big(\trionedcmb \\
& \qquad +\tilde \Z\Big(\trionedcIR\Big)\trionedcRE\Big)\,.
\end{align}
Here we introduced a doubly encircled vertex, which corresponds to a single encircled vertex with a further gluon's momentum nullified. Such a vertex vanishes in fact, and so for this reason the counterterm would not survive. Let us nevertheless continue with its evaluation to give an example of the procedure. The IR-counterterm can be evaluated by relating it to a UV-counterterm; this procedure has been used extensively in the $R^*$-literature for scalar diagrams and we straight forwardly extend it to the non-scalar case. Here we can do this as follows:
\begin{align}
&\tilde \Z\Big(\trionedcIR\Big)=-\Z\Big(\trionedcIRUV\Big)\nn\\
&=K(\trionedcIRUV)\,.
\end{align}
Having introduced the basic concepts let us now illustrate the procedure for the evaluation of a suitably differentiated and IR-rearranged two-loop diagram:
\begin{align}
&\Z\Big(\tritwoM\Big)=-K\Big(\tritwoM \nn\\
&+  \tilde\Z\Big(\tritwoMira \Big)\tritwoMuva \\
&+ \tilde\Z\Big(\tritwoMira \Big)\Z\Big(\tritwoMuva\Big)  \nn\\
&+ \Z\Big(\tritwouvb\Big)\tritwouvbrem\nn
\Big)
\end{align}
This diagram requires several counterterms, some of which we discard immediately due to scalelessness of the remaining/contracted graphs.
The second term on the right hand side captures a one-loop IR-subdivergence. The third term has the same IR-divergence with the remaining graph also giving rise to a UV-subdivergence. The last term corresponds to another UV subdivergence, which was originally of box-type.

\section{Calculation and results}

The actual calculation of the Feynman diagrams contributing to $\Gamma^{nb}_W$, which we first generate using QGRAF \cite{Nogueira:1991ex}, is done via two independent codes. The Levi-Cevita tensor appearing in the Weinberg operator is not strictly defined in $D$ dimensions and one must fix a scheme when encountering it within dimensional regularization. In the first code, written in Maple, we use the Larin~\cite{Larin:1993tq} scheme for the LCT appearing in the Feynman rules. The second code is written in \textsc{form}~\cite{Ruijl:2017dtg} and applies the 't Hooft-Veltman (HV) scheme \cite{tHooft:1972tcz}. The implementation of these schemes is further discussed in the appendix. Since the implementation of these two schemes results in rather different algorithms, obtaining a consistent result provides a powerful check. For the reduction to Master integrals both our implementations heavily rely on the \textsc{forcer} program~\cite{Ruijl:2017cxj}.

To calculate the two- and three-loop anomalous dimensions of the Weinberg operator one can extract $Z_W$ from  the correlator $\Gamma_W^{nb}$ for $n=3$-$6$. Ward identities ensure that the resulting anomalous dimension is independent of $n$. A smaller $n$ implies a lower number of diagrams to compute. However, this is counterbalanced by the fact that the $n$-background field correlator must be differentiated $6-n$ times with respect to external momenta in order for IRR to be applicable. Even though the Taylor expansion proliferates terms for the $n=3$ case, it is nevertheless the least computationally demanding and involves 250 diagrams. At the two-loop level we obtain the result 
\begin{align}
  \begin{split}
    &Z_{WW}(\mathrm{2-loop}) = \left(\frac{\alpha_s}{4\pi}\right)^2\left[C_A^2\left(-\frac{19}{24\epsilon^2}+\frac{119}{36\epsilon}\right)\right.\\
      &\left.-C_A\,n_f T_f\left(\frac{7}{3\epsilon^2}+\frac{4}{3\epsilon}\right)+\frac{3C_F n_f T_f}{\epsilon}+\frac{10n_f^2 T_f^2}{3\epsilon^2}\right]\,,
    \label{eq:ZWnlo}
   \end{split}
\end{align}
where $C_A$ is the adjoint Casimir, $C_A=N_c$ for the gauge group SU($N_c$), with $N_c$ the number of colors. 

We have checked our two-loop results in four ways. First, as discussed above we have applied two independent codes to the $n_f=0$ terms, obtaining the same result. Second, in the \textsc{form} code we investigated gauge invariance by performing the computation with a single power of the gauge parameter $\xi^1$ and
verified that it cancels. Third, we have extracted $Z_{WW}$ from the $n=4$ case. The evaluation of the associated 2389 diagrams leads to the same two-loop result for $Z_{WW}$. Finally, the $1/\eps^2$ poles can be determined from one-loop results, and we have verified that our results match the one-loop predictions.  

Using the \textsc{form} code, which is optimized for large expressions, we evaluated the $n_f=0$ three-loop correction to $Z_{WW}$. This required the computation of $6203$ diagrams involving up to $\mathcal O(10^9)$ terms in intermediate expressions. The total computation time came to 48 hours on a 24-core machine with 2.4~GHz Intel Xeon E5-2695v2 CPUs and 150~GB of memory, whereas the two-loop computation only took 20 minutes. We obtain the result
\begin{align}
  Z_{WW}(\mathrm{3-loop}) &= \left(\frac{\alpha_s}{4\pi}\right)^3\,\left\{C_A^3\left[\frac{779}{432\epsilon^3}-\frac{5389}{648\epsilon^2}-\frac{3203}{1944\epsilon}\right.\right.\nonumber\\
  &\left.\left.+\frac{44\zeta_3}{3\epsilon}\right]+\frac{d^{abcd}_Ad^{abcd}_A}{N_AC_A}\left[\frac{40}{3\epsilon}-\frac{352\zeta_3}{\epsilon}\right]\right\}.  \label{eq:ZWn2lo}
\end{align}
Here we encounter the Riemann zeta value, $\zeta_3\simeq1.202$, and the quartic Casimir $d^{abcd}_Ad^{abcd}_A/(N_AC_A)$, see Ref.~\cite{vanRitbergen:1998pn} for more details. For SU($N_c$) this becomes $d^{abcd}_Ad^{abcd}_A/ (N_AC_A) =N_c(N_c^2+36)/24$. We checked this result by verifying that the $1/\eps^2$ and $1/\eps^3$ poles match one- and two-loop predictions. Furthermore, the fact that $Z_{WW}$ is proportional to the three-gluon Feynman rule of the Weinberg operator is another non-trivial cross-check.

\section{Discussion}
The UV counterterm determines the anomalous dimension of the Weinberg
operator up to three loops
\begin{align}
  \label{eq:gammaW}
  \gamma_{WW}&=\frac{\alpha_s(\mu^2)}{4\pi}\left[\frac{C_A}{2}+\,2n_fT_f\right]\nonumber\\
  &+\left(\frac{\alpha_s(\mu^2)}{4\pi}\right)^2\left[\frac{119}{18}C_A^2+n_f T_f\left(6 C_F - \frac{8C_A}{3}\right)\right]\nonumber\\
  &+\left(\frac{\alpha_s(\mu^2)}{4\pi}\right)^3\bigg[C_A^3\Big(-\frac{3203}{648}+44\zeta_3\Big)\\
  &\qquad\quad+\frac{d^{abcd}_Ad^{abcd}_A}{N_AC_A}\Big(40-1056\zeta_3\Big)+\mathcal{O}\left( n_f \right)
  \bigg]\,,\nonumber
\end{align}
where we included the complete dependence on $n_f$, the number of active quark flavors, at one and two loops. Remarkably the two-loop $n_f$-dependence drops out since $C_F=4/3$ and $C_A=3$ for $N_c=3$. This is an accidental cancelation for $N_c=3$ but even for a larger number of colors the $n_f$ corrections are negligible due to the large prefactor of the $C_A^2$ term. 
It is interesting that $\zeta_3$ and the quartic group invariant enter at three loops, unlike
 the QCD beta function where they appear only at the four-loop order. However, to the best of our knowledge, there is nothing that forbids their appearance at lower order in the Weinberg operator. 
We notice that the coefficient of the quartic group invariant is directly proportional, up to a factor $2C_A/9$, to the same quartic group invariant appearing in the four-loop QCD beta function. While this could simply be a coincidence, it would be  interesting to see if similar patterns appear at higher loop orders. Finally, we have not calculated the $n_f$ corrections at three loops, but if a similar pattern appears as at two loops then neglecting their contributions would provide a good approximation.

In order to estimate the impact of the two- and three-loop contributions to the anomalous dimension we set
$C_A=3$, $C_F=\frac{4}{3}$ and $n_f=0$ in eq.~(\ref{eq:gammaW}) and we obtain the series
\be
\frac{8\pi\gamma_{WW}}{\alpha_s(\mu^2)C_A}= 1+3.15657\alpha_s -23.72872\alpha_s^2\,.
\ee
The next-to-next-to-leading-order (NNLO) coefficient can be decomposed as $23.72872=5.46537-\underline{29.19409}$
where the underlined number stems from the contribution of the quartic group invariant, which is responsible for the large negative correction.
The size of the coefficients increases drastically with the loop order, and undermines the convergence of the $\alpha_s$ expansion, unless  large cancellations occur in the $n_f$-dependent pieces of the three-loop anomalous dimensions that are not computed in this work. 

As an example, we calculate the evolution of $C_W$, determined by
\begin{equation}
  \mu^2\frac{d C_W(\mu)}{d\mu^2} = \gamma_{WW}\,C_W(\mu)\,,
  \label{eq:CWrun}
\end{equation}
from a specific high-energy scale $\mu_H =1$ TeV, where we assume
$C_W(\mu_H)=1$, to various low-energy scales.
We use $\alpha_s(M_Z) = 0.118$ and $M_Z=91.2\,{\mathrm{GeV}}$
\cite{Tanabashi:2018oca}, and apply the QCD beta function at two
\cite{Caswell:1974gg,Jones:1974mm,Egorian:1978zx} and three loops
\cite{Tarasov:1980au,Larin:1993tp}. When evaluating the beta function we 
adjust the number of fermions $n_f$ at the top, bottom and charm
thresholds
\begin{align}
  m_t(m_t) &= 160\,\mathrm{GeV}\,,\quad m_b(m_b) = 4.18\,\mathrm{GeV}\,,\nonumber\\
  m_c(m_c) &= 1.28\, \mathrm{GeV}\,.
\end{align}
The Wilson coefficient at the different energy scales is given in
Tab. \ref{Tab:CW}, where we kept the LO and NLO $n_f$ dependence. Around and above the electroweak scale, $\alpha_s$ is sufficiently small such that NLO and NNLO corrections are suppressed. For $\mu \leq 100$ GeV, NNLO corrections are as large, or larger, than NLO corrections. At lower energies higher orders become relevant, and in particular for $\mu =1$ GeV, the scale where the Weinberg operator is often matched to hadronic quantities, the NLO and NNLO correction are $-21\%$ and $+33\%$ of the LO result, respectively. The total result, however, is not far from the LO result due to cancellations between NLO and NNLO corrections. The lack of convergence is worrying and warrants a four-loop calculation.

\begin{table}[h!]
\begin{tabular}{ | c | >{\centering}m{1cm} | >{\centering}m{1cm} | c | }
    \hline
    $\mu\left[\text{GeV}\right]$& LO & NLO & NNLO \\ \hline
    100 & 0.76 & 0.75 & 0.76 \\\hline
    5 & 0.48 & 0.44 & 0.48 \\\hline
    2 & 0.39 & 0.34 & 0.40 \\\hline
    1 & 0.33 & 0.26 & 0.37 \\\hline
\end{tabular}
\caption{\label{Tab:CW} Evolution factors at different perturbative order that relate $C_W(\mu)$ to $C_W(1\,\mathrm{TeV})$.}
\end{table}

The main phenomenological impact of a nonzero Weinberg operator is its contribution to the neutron EDM $d_n$. The QCD matrix element
connecting $d_n$ to $C_W$ is difficult to calculate, but future lattice-QCD calculations might be up to the task \cite{Rizik:2018lrz,Gupta:2019fex,Cirigliano:2019jig}. Two techniques have been used to estimate the matrix element. A QCD sum-rule estimate \cite{Demir:2002gg, Haisch:2019bml} gives $d_n = (25\pm12)\,\mathrm{MeV} \,e\,C_W(1\,{\mathrm{GeV}})$.   Another technique that is often applied is Naive Dimensional Analysis (NDA) \cite{NDA}. NDA predicts \cite{Weinberg:1989dx, deVries:2010ah}
\begin{equation}
  |d_n|\simeq e \frac{\Lambda_{\mathrm{\chi}}}{4\pi}\,C_W(\mu_{\mathrm{match}}),
  \label{eq:NDA}
\end{equation}
where $\Lambda_{\mathrm{\chi}}\simeq 1.2$ GeV denotes the
chiral-symmetry-breaking scale and $\mu_{\mathrm{match}}$ a matching
scale at hadronic energies. Unlike, the QCD sum rules calculation the NDA estimate is sensitive to the evolution of the Weinberg operator to the low hadronic matching scale. Which scale to pick is unclear, but typically the scale where $\alpha_s(\mu_{\mathrm{match}})=2\pi/3$ is applied as suggested by Weinberg \cite{Weinberg:1989dx}. Using one-loop evolution this leads to $|d_n^{\mathrm{LO}}|\simeq  e\,40\,\mathrm{MeV}\,C_W^{\mathrm{LO}}(1\,{\mathrm{GeV}})$ in reasonable agreement with QCD sum rules as pointed out in \cite{Demir:2002gg}. However, the large NNLO corrections significantly affect the running in the non-perturbative regime and lead to much larger estimates $|d_n^{\mathrm{NNLO}}|\simeq  e\,1\,\mathrm{GeV}\;C_W^{\mathrm{LO}}(1\,{\mathrm{GeV}})$, indicating that NDA estimates are not stable. We therefore recommend the use of QCD sum rules \cite{Demir:2002gg, Haisch:2019bml}, which only depend on the evolution to the perturbative scale $\mu_L \simeq 1$ GeV. Ideally, these calculations are replaced by lattice-QCD results in the future \cite{Rizik:2018lrz,Gupta:2019fex,Cirigliano:2019jig,Rizik:2020naq}. 

Using the conservative QCD sum rule expression $d_n = 13\,\mathrm{MeV} \,e\,C_W(1\,\mathrm{GeV})$, and our result for the anomalous dimension, the current neutron EDM limit, $d_n < 1.8 \cdot 10^{-13}$ e fm \cite{Abel:2020gbr} can be used to constrain the Weinberg operator. We write $C_W(\Lambda) = d_W(\Lambda)/\Lambda^2$ where $d_W(\Lambda)$ is a dimensionless constant. For $\Lambda =1$ TeV, we obtain the constraint $d_W(1\,\mathrm{TeV}) < \{2.1,\,2.7,\,1.9\} \cdot 10^{-4}$ where the results in brackets are obtained with the LO, NLO, and N${}^2$LO anomalous dimension, respectively. The limits are rather stringent, in particular in comparison with limits on the CP-even counterpart of the Weinberg operator
\begin{equation}
  \mathcal L_{6} = \frac{c_G}{\Lambda^2} f^{abc} G^{a,\,\rho}_{\nu}G^{a,\,\nu}_{\lambda}G^{c\,\lambda}_{\rho}\,,
  \label{def:weinberg2}
\end{equation}
 that is constrained at the percent level $c_G (1\,\mathrm{TeV}) \leq 4 \cdot 10^{-2}$ from an analysis of multi-jet production at the LHC \cite{Krauss:2016ely}.

\section{Conclusion}
In this paper we have reported a new method to calculate higher-order QCD anomalous dimensions of SMEFT operators in a highly automated manner. To develop and test the method we focused on one particular operator, the CP-violating gluonic Weinberg operator, whose anomalous dimension is hard to calculate, even at one loop \cite{Weinberg:1989dx,BraatenPRL,Braaten:1990zt,Chang:1990dja}. Due to its CP-violating nature this operator does not mix into lower-dimensional operators containing just gauge fields. By also applying the background field method, we avoid complications associated to operator mixing. We extracted the two-loop anomalous dimension by calculating 250 diagrams contributing to the three background-gluon vertex. We verified our result by extracting the same anomalous dimension of the 2389 diagrams contributing to the four background-gluon vertex, as predicted by gauge invariance. 
Finally, due to the automated nature of the framework we were able to immediately calculate the three-loop anomalous dimension at $n_f=0$ by evaluating 6203 diagrams. 

We found a sizeable positive two-loop correction to the anomalous dimension, which turns out to be independent on $n_f$, the number of flavors, due to an accidental cancellation for $N_c=3$. Even at $N_c \neq 3$, the $n_f$-dependent corrections are negligible. We proceeded to calculate the three-loop correction in $n_f=0$ limit and found a negative contribution, which is sufficiently large to threaten the perturbative convergence, as the two- and three-loop evolution factors almost cancel. 
 The lack of convergence motivates a calculation of the four-loop anomalous dimension and the missing three-loop $n_f$-corrections. 

The $R^*$-operation combined with the background field method provides a powerful framework for higher-order loop calculations, and our calculation can be extended into several directions without too much additional effort. In fact, our setup works up to five loops. Currently the only hurdle is computing time. Beyond five-loops --- should the need for such corrections ever arise --- there exists neither a complete basis of master integrals nor a suitable reduction onto such a basis. However, the $R^*$-operation itself is valid to all loop orders. 

So far, we have only performed higher-order loop calculations for operator without external quarks, and the framework must be extended to renormalize SMEFT operators containing quark fields. There are no inherent additional complications associated to the inclusion of quarks, 
but a consistent treatment of $\gamma^5$ in the $R^*$-method needs to be developed. This is left to future work. Here we have taken a first big step by presenting methods for the consistent use of LCTs beyond one loop in both the L and HV schemes.
Once developed, we can calculate the (so far unknown) higher-loop mixing of the Weinberg operator into the quark electric and chromo-electric dipole moments, and CP-odd four-quark operators. It would also be interesting to compute four-loop corrections to the anomalous dimension, to see whether the trend of large coefficients  continues and whether other surprising relations, as the one found for the quartic group invariants, to the QCD beta function appear. 

While this work focused on the Weinberg operator, this is by no means an inherent limitation of the method. We envision calculations of higher-order anomalous dimensions of a much larger class of SMEFT operators. Isolating the footprints of SMEFT operators left behind at the LHC, or future colliders, from SM contributions is an active field of research \cite{Maltoni:2016yxb,Hartland:2019bjb}. Higher-order QCD corrections to SMEFT contributions can be sizeable, as shown here, and are important to disentangle SMEFT operators \cite{Alioli:2018ljm}. Higher-loop anomalous dimensions will further reduce theoretical uncertainties and make the theoretical framework of the SMEFT more robust.

\section*{Acknowledgements}
We are grateful for conversations with Emanuele Mereghetti, Sven Moch, Giovanni Marco Pruna, Adam Ritz, Peter Stoffer, Jos Vermaseren and Stefano di Vita.
The research of FH is supported by the NWO Vidi grant 680-47-551. JdV is supported by the 
RHIC Physics Fellow Program of the RIKEN BNL Research Center.
GF received support from the ERC Advanced Grant no. 320651, ``HEPGAME''.
The research of BR is supported by the ERC Advanced Grant no. 694712, ``PertQCD''.

\newpage

\appendix

\section{Levi-Civita Symbols and the $R^*$-method}\label{sec:levi_scheme}
We define the LCTs in the respective schemes by $\epsv_{HV}^{\mu_1 \ldots \mu_4}$ and $\epsv_{L}^{\mu_1 \ldots \mu_4}$. While $\epsv_{HV}$, associated to the 't Hooft-Veltman (HV) scheme~\cite{tHooft:1972tcz}, lives in the 4-dimensional subspace, $\epsv_{L}$, associated to the Larin scheme~\cite{Larin:1993tq}, has components in the full $D$-dimensional space.
As such, we can freely commute $\epsv_{HV}$ with the pole operation $K$ which appears in the $R^*$-counterterm operation:
\be
\label{eq:epscommute}
K(\epsv_{HV}^{\mu_1 \ldots \mu_4} F_{\mu_1 \ldots \mu_4 \ldots})=\epsv_{HV}^{\mu_1 \ldots \mu_4}K( F_{\mu_1 \ldots \mu_4 \ldots})\,.
\ee

A simple, although inefficient, procedure to use the local $R^*$-operation in the HV scheme is to commute $\epsv_{HV}$ out of all the potentially nested $K$s and then to apply a standard tensor-reduction on the remaining tensor as described in Ref.~\cite{Ruijl:2018poj}. In the L-scheme eq.~\eqref{eq:epscommute} does not hold and 
we apply a different procedure. To compute the expression  
\be 
K \left( F_\epsv^{\nu_1 \ldots \nu_n} \right)=K \left(\epsv_{L}^{\mu_1 \ldots \mu_4} F_{\mu_1 \ldots \mu_4}^{\nu_1 \ldots \nu_n}\right)\,,
\ee 
we first tensor-reduce the object $F_\epsv^{\nu_1 \ldots \nu_n}$, which (assuming that the object is superficially log-divergent) can be reduced in terms of $\epsv_{L}^{\mu_1 \ldots \mu_4}$ and metric tensors $g^{\mu_1\mu_2}$. We thus write:
\be
K \left( F_\epsv^{\nu_1 \ldots \nu_n} \right )= \sum_{\sigma} T^{\nu_1 \ldots \nu_n}_\sigma K\left( F_\epsv^\sigma \right)\,,
\ee
where 
\be
T^{\nu_1 \ldots \nu_n}_\sigma=\epsv_L^{\nu_{\sigma(1)}\ldots\nu_{\sigma(4)}}g^{\nu_{\sigma(5)}\nu_{\sigma(6)}} \cdots g^{\nu_{\sigma(n-1)}\nu_{\sigma(n)}}\,,
\ee
and the sum goes over all permutations of the indices which do not leave the tensor structure invariant. The coefficients $F_\epsv^\sigma$ are defined 
\be
 F_\epsv^\sigma=P^\sigma_{\nu_1 \ldots \nu_n}F_\epsv^{\nu_1 \ldots \nu_n}\,, 
\ee
where the projector $P^\sigma$ satisfies 
\be 
P^\sigma_{\nu_1 \ldots \nu_n}T^{\nu_1 \ldots \nu_n}_\tau=\delta_{\sigma\tau}\,,
\ee 
and $\delta_{\sigma\tau}$ is a Kronecker-delta which yields $1$ if the two permutations $\tau$ and $\sigma$ are identical and $0$ otherwise.
The $P^\sigma$ themselves can be constructed as linear combinations of $T_\sigma$s. 
Using the identity
\be
\epsv_L^{\nu_{1} \ldots \nu_{4}}\epsv_L^{\mu_{1} \ldots \mu_{4}}=
\det
\begin{pmatrix} 
g^{\mu_1\nu_1} & g^{\mu_1\nu_2} & \ldots &  g^{\mu_1\nu_4}\\ 
g^{\mu_2\nu_1} & g^{\mu_2\nu_2} & \ldots &  g^{\mu_2\nu_4}\\ 
\ldots &\ldots&\ldots&\ldots\\
g^{\mu_4\nu_1} & g^{\mu_4\nu_2} & \ldots &  g^{\mu_4\nu_4}\\ 
\end{pmatrix}
\ee
products of two LCTs in the L-scheme can always be evaluated in terms of D-dimensional metric tensors. 
The scalar functions, $F_\epsv^\sigma$ are thus functions of D-dimensional scalar products only, 
and therefore may never contain any LCTs.

\bibliographystyle{utphys}%
\bibliography{refs}

\end{document}